%% file: tarallo.tex
\documentclass[runningheads]{llncs}

\usepackage{graphicx}
\usepackage{amssymb}
\usepackage{amsmath}
\usepackage{svg}
\usepackage{algpseudocode}
\usepackage[Algorithm]{algorithm}
\usepackage{todonotes}
\usepackage{paralist}
\usepackage{booktabs}
\usepackage[nolist]{acronym}
\usepackage{xspace}
\usepackage[american]{babel}
\usepackage{tcolorbox}
\usepackage{multirow}

\definecolor{apricot}{HTML}{FBB982}

\begin{document}

\title{Tarallo: Evading Behavioral Malware Detectors in the Problem Space}

\newcommand{\mypar}[1]{\smallskip\noindent\textbf{#1.}}

\newcommand{\myack}[1]{\bigskip\noindent{\bfseries\fontsize{10pt}{12pt}\selectfont #1}\xspace}

\input{authors}
\input{acronyms}

\subtitle{\normalfont\scriptsize This paper has been accepted at the International Conference on Detection of Intrusions and Malware, and Vulnerability Assessment (DIMVA '24). Cham: Springer Nature Switzerland, 2024.}

\maketitle

\begin{abstract}
Machine learning algorithms can effectively classify malware through dynamic behavior but are susceptible to \textit{adversarial attacks}. Existing attacks, however, often fail to find an effective solution in \textit{both} the feature and problem spaces.
This issue arises from not addressing the intrinsic nondeterministic nature of malware, namely executing the same sample multiple times may yield significantly different behaviors. Hence, the perturbations computed for a specific behavior may be ineffective for others observed in subsequent executions. In this paper, we show how an attacker can augment their chance of success by leveraging a new and more efficient feature space algorithm for sequential data, which we have named \acl{PS-FGSM}, and by adopting two problem space strategies specially tailored to address nondeterminism in the problem space. We implement our novel algorithm and attack strategies in \textit{Tarallo}, an end-to-end adversarial framework that significantly outperforms previous works in both white and black-box scenarios. 
Our preliminary analysis in a sandboxed environment and against two \ac{RNN}-based malware detectors, shows that
Tarallo achieves a success rate up to 99\% on both feature and problem space attacks while significantly minimizing the number of modifications required for misclassification.
\end{abstract}


\input{sections/introduction}
\input{sections/related_work}
\input{sections/preliminaries}
\input{sections/methodology}

\input{sections/experiments}

\input{sections/conclusions}

\myack{Acknowledgements}
\fontsize{9pt}{10pt}\selectfont
This study was carried out within the MICS (Made in Italy – Circular and Sustainable) Extended Partnership and received funding from Next-Generation EU (Italian PNRR – M4 C2, Invest 1.3 – D.D. 1551.11-10-2022, PE00000004). CUP MICS D43C22003120001. Mario D’Onghia acknowledges support from TIM S.p.A. through the PhD scholarship.

\bibliographystyle{splncs04}
\bibliography{bibliography.bib}

\end{document}

%% file: authors.tex
\author{  
    Gabriele Digregorio\inst{1}\and 
    Salvatore Maccarrone\inst{1} 
    \and
    Mario D'Onghia\inst{1} 
    \and
    Luigi Gallo\inst{2} 
    \and
    Michele Carminati\inst{1} 
    \and
    Mario Polino\inst{1} 
    \and
    Stefano Zanero\inst{1} 
}

\authorrunning{Digregorio et al.}
\titlerunning{Tarallo: Evading Behavioral Malware Detectors in the Problem Space}

\institute{
    Politecnico di Milano, Milan, Italy
    \email{\{gabriele.digregorio,mario.donghia,michele.carminati,
    mario.polino,stefano.zanero\}@polimi.it}\\
    \email{salvatore.maccarrone@mail.polimi.it}\\
    \and
    Cybersecurity Lab, TIM S.p.A., Turin, Italy\\
    \email{luigi1.gallo@telecomitalia.it}
}

%% file: acronyms.tex
\begin{acronym}
    \acro{ML}{Machine Learning}
    \acro{DL}{Deep Learning}
    \acro{AML}{Adversarial Machine Learning}
    \acro{DNN}{Deep Neural Network}
    \acro{RNN}{Recurrent Neural Network}
    \acro{LSTM}{Long Short-Term Memory}
    \acro{API}{Application Programming Interface}
    \acro{GAN}{Generative Adversarial Network}
    \acro{FGSM}{Fast Gradient Sign Method}
    \acro{PS-FGSM}{\textit{Position Sensitive - Fast Gradient Sign Method}}
    \acro{PE}{Portable Executable}
    \acro{LKB}{Longest Known Behavior}
    \acro{BCO}{Behavior Cascade Optimization}
    \acro{TPR}{True Positive Rate}
    \acro{FPR}{False Positive Rate}
    \acro{CNN}{Convolutional Neural Network}
    \acro{AUC}{Area Under the Curve}
    \acro{IDS}{Intrusion Detection System}
\end{acronym}

%% file: sections/introduction.tex
\section{Introduction}\label{sec:introduction}

Malware detectors rely on \textit{software analysis techniques} to identify malicious programs. These techniques are grouped into two main categories: \textit{static} and \textit{dynamic}. Static analysis techniques examine the code without running the program, such as strings or IP addresses within the code~\cite{hassen2017malware}. While providing a fast way to analyze software, static analysis falls short against obfuscated samples~\cite{d2022apicula,survey_malware_analysis,galloro2022systematical,you2010malware}. On the other hand, \textit{dynamic analysis} gathers information while the program is running, capturing its \textit{dynamic behavior}. This includes system calls, network traffic, and memory usage. One commonly employed dynamic feature is the sequence of executed \ac{API} calls, which can effectively represent the program functions and goals. 

\ac{ML}, and particularly \ac{DL}, can be employed to detect or classify malware, leveraging both static~\cite{malconv,ember,drebin} and dynamic~\cite{clustering_behaviors,zhang2020dynamic,or2019dynamic} features. They represent a more effective alternative to traditional approaches focusing on pattern identification (e.g., byte signatures) or heuristics due to their well-understood generalization capabilities~\cite{moser2007limits,catak2020deep}. Nonetheless, \ac{ML} algorithms are susceptible to \textit{adversarial} attacks, which consist in forcing an \ac{ML} model to misclassify specially \textit{perturbed} samples~\cite{goodfellow2014explaining}. These attacks may heavily impact the reliability of \ac{ML} for cybersecurity applications, forcing their advocates to preemptively research and address vulnerabilities. A substantial body of work in adversarial attacks against static analysis and \ac{ML}-based malware detectors exists in the literature~\cite{kreuk2018deceiving,kolosnjaji2018adversarial,suciu2019exploring,malware_makeover,lookinout}. Fewer works have addressed the security of classifiers employing dynamically extracted features. 
Of great relevance are the works by Rosenberg et al.~\cite{rosenberg2018generic,rosenberg2020query}, which introduced an adversarial framework that can target a variety of \ac{ML}-based malware detectors that employ \ac{API} call sequences. The authors also discussed and showed how to modify a program's dynamic behavior without \textit{breaking} its functionality, an essential requirement for the attack to be considered successful, formally known as preserved semantics~\cite{intriguing_properties_adv_samples}.
However, these works do not explicitly discuss or evaluate the real impact of their attacks in the \textit{problem space}. Our preliminary experiments show that only ``attacking'' the final \ac{ML} classifier does not suffice to evade the \textit{whole} detection system. This is due to the nondeterminism and probabilistic nature of malware behavior. Hence, computing an adversarial perturbation after observing a specific behavior may not be enough to guarantee a successful attack in subsequent executions. Furthermore, as shown in Section~\ref{sec:experiments}, the original feature space attack proposed by Rosenberg et al.~\cite{rosenberg2018generic} requires a very large number of additional \ac{API} calls to evade newer and more sophisticated classifiers such as~\cite{li2022novel,zhang2020dynamic}.
This may affect both the stealthiness of the attack as well as its capacity to deliver its original and intended behavior.

Our main contribution consists in designing an \ac{AML} attack that is effective both in the feature \textit{and} problem spaces while targeting systems that employ sophisticated \ac{DL} algorithms such as~\cite{li2022novel,zhang2020dynamic}. First, we propose \acl{PS-FGSM}, a new adversarial attack specific to sequential data, able to produce evasive behaviors with minimal perturbations. Then, we address the problem space limitations through two variants: with the first, the attacker executes several times the malware sample, recording all behaviors, and then selects the longest. The second approach consists of performing the attack while considering multiple behaviors at once.

We preliminary evaluate our attacks through an array of experiments against two state-of-the-art \acfp{RNN}-based malware detectors, measuring the attack success in the \textit{problem space}~\cite{intriguing_properties_adv_samples} in both white- and black-box scenarios. We confirm that after modification and re-execution in the sandbox, the adversarial samples can still evade detection. Our more stringent criterion contrasts with those adopted in previous works~\cite{rosenberg2018generic,rosenberg2020query}, which solely focused on evaluating the evasion rate in the feature space. 
Tarallo achieved a success rate of $99\%$ in the feature space.  Additionally, it significantly reduced the number of injections required to achieve misclassification, with an average of $27$ additional API calls required, a value significantly lower than what is required by current state-of-the-art approaches against weaker models. In the problem space experiments, our method attains a success rate of up to $99\%$.
While securing high evasion rates, our approach also maintains the original malware functionality. In $89\%$ of the cases, the modified malware preserves its original behavior, thus ensuring its capability to execute its original malicious functions while evading detection. Overall, the results from our experimental evaluations highlight the effectiveness and efficiency of our approach, particularly when compared to state-of-the-art attacks. 

The main contributions of this research are summarized below:
\begin{itemize}
    \item We introduce \acl{PS-FGSM}, a novel \ac{AML} attack algorithm that targets discrete sequences, which, against \ac{RNN} models, proves to be more effective and efficient than state-of-the-art approaches.

    \item We design two strategies to compute an optimal \ac{API} call sequence that aims to evade detection in the problem space, addressing the nondeterminism between executions. 
    
    \item We propose a less invasive code-modification strategy that preserves the original functionality of modified malware samples.
    
    \item We present Tarallo, an end-to-end framework that modifies the apparent dynamic behavior of malware, revealing vulnerabilities in state-of-the-art \ac{RNN}-based detection systems. We make Tarallo publicly available to ensure reproducibility and encourage further research on the topic\footnote{https://github.com/necst/Tarallo}.
\end{itemize}

%% file: sections/related_work.tex
\section{Related Work}\label{sec:related_work}

\mypar{Mimicry Attacks}
In the original formulation~\cite{wagner2000intrusion}, mimicry attacks aimed to camouflage malicious activities by emulating legitimate goodware behavior. A follow-up work refined this strategy, enabling to fool \acp{IDS} that monitor system calls by adding ones that have no effect or swapping some for equivalent system calls~\cite{wagner2002mimicry}.
Nevertheless, mimicry attacks, as well as the related countermeasures (e.g.,~\cite{giffin2006automated}), targeted \acp{IDS} relying on simpler anomaly detection techniques (e.g., by comparing recorded behaviors against known benign ones and modeling behaviors as automata~\cite{warrender1999detecting,somayaji2000automated,why6}) than the one based on \ac{ML} and \ac{DL} targeted in this work.  

\mypar{DL-based Behavioral Malware Detection}
This section presents a brief discussion of \ac{DL} algorithms for malware detection employing dynamically extracted \ac{API} call sequences as features. In particular, we focus on the use of \acp{RNN} to process long sequences of \ac{API} calls, as those are the targets of the attacks presented in this paper. We refer the readers interested in a more in-depth analysis of the topic, including the use of traditional \ac{ML} as in~\cite{uppal2014malware,tian2010differentiating,fang2017new}, to~\cite{berman2019survey}.  

In \cite{zhang2020dynamic}, the authors presented a feature representation for arguments of \ac{API} calls, in contrast with most existing works that focus primarily on \ac{API} names. They designed a \ac{DNN} architecture tailored to process these features, which are encoded using a hashing trick method. 
Samples are executed in a sandbox that records the sequences of API calls. To capture the relationships between these API calls, the model incorporates gated-\acp{CNN}, batch normalization layers, a bidirectional \ac{LSTM}~\cite{agrawal2018neural}, a global max-pooling layer, and a dense layer. This configuration achieved an accuracy of $\approx 95\%$  and a recall of $\approx 71\%$.

In \cite{li2022novel}, the authors introduced a \ac{DNN} for dynamic malware detection based on intrinsic features extracted from \ac{API} call sequences. Similar to~\cite{zhang2020dynamic}, each malware sample is executed in a sandbox, where all the \ac{API} call sequences are recorded. However, instead of using arguments to encapsulate the semantic information of each \ac{API} call, their proposed model considers three distinct attributes: the category, the action, and the operation object associated with it. Let us consider the \ac{API} \textit{RegCreateKeyExW}: here, \textit{Create} is the action, \textit{RegKeyEx} the operation object, and \textit{registry} the category. The architecture of the model consists of embedding layers, multi-layer \acp{CNN}, a bidirectional \ac{LSTM}, and dense layers. This architecture achieved an accuracy and recall of 97\% and 98\%, respectively.

\mypar{Adversarial Machine Learning}
Research in \ac{AML} has focused on inducing misclassification in \ac{ML} classifiers at inference time, particularly in the image domain~\cite{eykholt2018robust,machado2021adversarial,goodfellow2014explaining}. 
However, traditional \ac{AML} attacks for images (e.g., FGSM~\cite{goodfellow2014explaining}) cannot be directly applied to malware detectors due to specific problem-space constraints (i.e., functionality preservation) and the discrete and sequential nature of API call sequences.

There is a growing body of research on perturbing Windows \ac{PE} files; in particular, attacks on static features have been extensively explored in the literature. These attacks involve modifying the headers~\cite{rosenberg2020generating,rosenberg2020bypassing} or the raw bytes of PE files~\cite{lookinout,kolosnjaji2018adversarial,suciu2019exploring,malware_makeover}. Fewer works have instead targeted dynamic and machine learning-based malware detectors~\cite{ming2017impeding,hu2023generating,hu2017black,rosenberg2018generic,rosenberg2020query}. While Ming et al. focused on evading a particular detection strategy based on the \textit{system call dependency graph}~\cite{ming2017impeding}, we tackle a more general and powerful detection approach that looks at the actual dynamic behaviors (namely, the \ac{API} call sequence) as discriminating features. 

Most closely related to ours are the works by Rosenberg et al.~\cite{rosenberg2018generic,rosenberg2020query}, which propose an adversarial framework for deceiving \ac{ML} detectors that work on sequences of dynamically recorded \ac{API} calls, including \acp{RNN} and feed-forward \acp{DNN}. Furthermore, they included a tool that modifies malware samples to reflect the required changes in their dynamic behavior.
In contrast to~\cite{rosenberg2018generic}, our experimental evaluation considers state-of-the-art \ac{DL}-based behavioral malware detectors. We show that the method from~\cite{rosenberg2018generic} is less effective and efficient against these advanced models, while our strategy proves both highly effective and efficient in evading them. Additionally, Rosenberg et al.~\cite{rosenberg2018generic} assess their attack effectiveness solely in the feature space, deeming it successful if the adversarial algorithm can transform a malicious \ac{API} call sequence into a variant misclassified as benign by the target detector. In contrast, our study adopts a more complex and realistic approach. An attack is deemed successful if it can alter a malicious sample in a manner that, \textit{upon re-execution}, it produces an \ac{API} call sequence misclassified as benign by the target model. Our method addresses the significant challenge of varying \ac{API} call sequences that result from different executions of the same malware sample, an aspect overlooked in~\cite{rosenberg2018generic}.
Lastly, in~\cite{rosenberg2018generic} the authors suggest using a wrapper as proxy code, serving as an intermediary between the malware and the DLLs executing the \ac{API} calls. Our method directly
modifies the malware bytecode to intercept \ac{API} calls. This alteration of the bytecode itself removes the necessity for an extra proxy layer, making the attack more realistic and potentially lowering the likelihood of detecting the modification.
Rosenberg et al. also introduced a more efficient black-box attack, necessitating fewer queries to the target model~\cite{rosenberg2020query}. This method employs a Generative Adversarial Network~\cite{creswell2018generative} to mimic genuine benign \ac{API} call sequences. The authors evaluated the efficacy of this attack with varying query budgets. While it outperforms their earlier method~\cite{rosenberg2018generic} under a limited query budget, their findings also indicate that a gradient-based attack becomes more effective with a larger budget. In contrast, our adversarial algorithm not only surpasses the gradient-based attack from~\cite{rosenberg2018generic} in efficiency and query requirement but also remains effective even with minimal or no knowledge of the target model.

%% file: sections/preliminaries.tex
\section{Background \& Threat Modeling}

\mypar{Adversarial Malware}\label{sec:preliminaries}
Based on the work by Pierazzi et al.~\cite{intriguing_properties_adv_samples}, we formally define the notions employed in this paper.
Consider a set of executable programs denoted as $E$. Let $F$ represent the feature space in which our model operates. We define $g$ as the feature extraction function that maps an executable to a specific point within the feature space, i.e., \( g:E \rightarrow F\). The readers must notice that $g$ is generally not a surjective function, as it may be generally not possible to identify an executable corresponding to a certain point in the feature space~\cite{goodfellow2014explaining}. Let $M$ represent a machine learning classifier, which accepts as input a point in the feature space representing a sequence of API calls. $M$ outputs a prediction from the predefined label set $L = \{malware, goodware\}$, i.e., $  M:F \rightarrow L$. Given an executable \( e \) belonging to the set \( E \) and its corresponding representation in the feature space $x_e\in F$, we define the adversarial sample \( \hat{x_e} \) for the model \( M \) as 
$\hat{x_e} \triangleq x_e + \delta$ where  $ \delta \in F $ is the adversarial perturbation, such that $ M $ classifies correctly $ x_e $ while misclassifies $\hat{x_e}$. 


Crafting adversarial samples in the malware problem space presents significant challenges. In contrast to computer vision, where attackers aim to introduce subtle and imperceptible perturbations to images, adversarial malware must be valid and executable programs. This requirement limits the range of possible points in the feature space that can be considered, as not all points correspond to valid and executable programs. Furthermore, the original functionality of the program must be preserved for the attack to be considered successful. This further restricts the extent of possible modifications that an attacker can apply. An attacker can only add extra API calls to the original sequence; in fact, deleting or modifying existing ones would compromise its original functionality. 

\mypar{Threat Model}
We examine two different scenarios based on the attacker's knowledge of the target model. In the white-box scenario, the attacker has complete access to its internal parameters. In this case, the model used as the target in the white-box attack directly serves as an oracle. 
In the black-box scenario, the attacker has no access to the target model architecture or parameters. Instead, the attacker relies on a surrogate oracle, which may have a different structure than the target detector. The only assumption in our experimental evaluation is that the target model employs \ac{API} call sequences as features.

%% file: sections/methodology.tex
\section{Tarallo}\label{sec:approach}

\begin{figure}[t]
\centering
\includegraphics[width=0.7\textwidth]{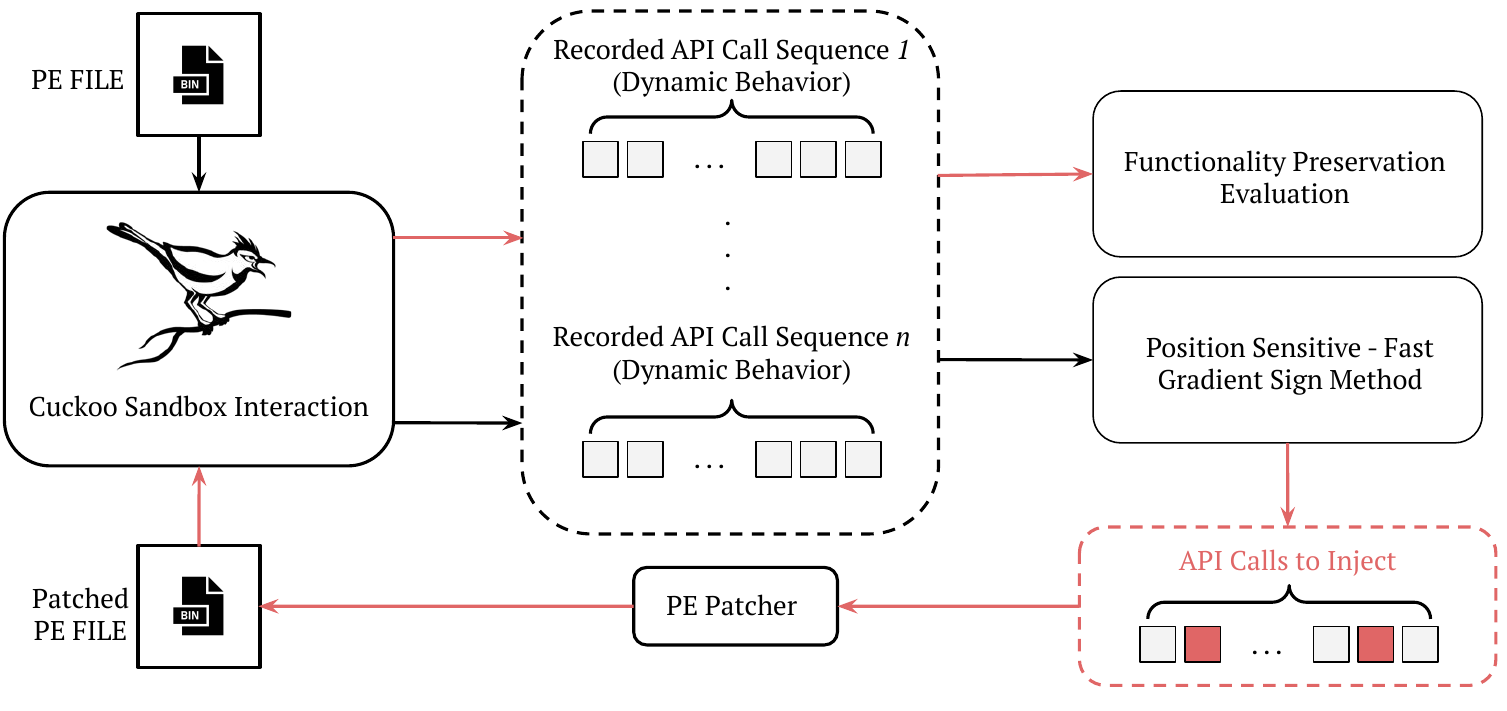}
\caption{The workflow of the Tarallo framework.}
\label{fig:Tarallo Framework}
\vspace{-0.5 cm}
\end{figure}

The Tarallo framework operates in an end-to-end manner, to modify the apparent dynamic behavior of malware. It comprises three key components, the first being a Cuckoo sandbox~\cite{cucuzzo} controller that automates the submission of malware samples and the retrieval of the corresponding reports. The second component is responsible for computing the \textit{adversarial} \ac{API} call sequence and consists of the \ac{PS-FGSM} algorithm and the two problem space strategies to maximize the evasion probability. Lastly, it comprises a \textit{PE patcher} that parses and modifies a PE file to make it reflect the evasive behavior computed by the second component. 
Figure~\ref{fig:Tarallo Framework} depicts Tarallo's workflow. To transform a given malware sample into a functioning adversarial sample, it performs the following steps:
\begin{inparaenum}
    \item It executes the sample multiple times in a controlled environment (sandbox) and records the dynamic behavior of each execution as a sequence of \ac{API} calls.
    \item It runs the \ac{PS-FGSM} to manipulate the original \ac{API} call sequence and generate a modified version that can evade the target machine learning model.
    \item It modifies the original malware sample to alter its apparent dynamic behavior so that it closely resembles the evasive behavior produced by the adversarial machine learning attack.
    \item It executes the modified sample multiple times in the sandbox, recording its dynamic behavior. 
    \item It lastly evaluates the attack success by examining the effect on the target machine learning model, while also ensuring the preservation of its functionality.
\end{inparaenum}

\subsection{Position Sensitive - Fast Gradient Sign Method}\label{ss:Adversarial ML Attack}%

The main contribution of this work is a novel \ac{AML} attack designed to target \ac{ML} models that rely on discrete and sequential data. We call this attack \acl{PS-FGSM} as it builds upon the \textit{\acl{FGSM}} introduced by Goodfellow et al.~\cite{goodfellow2014explaining}. Specific to malware detection, we employ \ac{PS-FGSM} against models that employ \ac{API} call sequences as input features. The algorithm relies on an oracle with known architecture and parameters. It can either work in a white-box mode, in which the oracle is the target model, or in a full black-box manner.
As a variation of the \acl{FGSM}~\cite{papernot2016crafting,goodfellow2014explaining}, the algorithm we present in this work relies on the \textit{sign} of the \textit{inverse} Jacobian of the model's output with respect to its input to minimize the probability of the sample to be classified as belonging to its real class, in this case ``malware.'' As a reminder, the Jacobian matrix represents all the partial derivatives of a function. Since we are considering the \textit{inverse} Jacobian, we are computing the \textit{gradients} of the model from the output to the input.

The \ac{PS-FGSM} repetitively selects the \textit{best position} within a given sub-sequence of consecutive \ac{API} calls (henceforth called \textit{windows}) and adds an extra \ac{API} call at that position. The algorithm iteratively attacks windows instead of whole sequences because the targeted models are most often \acp{RNN} which process one window per time. We take inspiration for the \ac{PS-FGSM} from the algorithm originally introduced by Rosenberg et al.~\cite{rosenberg2018generic}, significantly improving its effectiveness. In particular, we found a way to explicitly compute the position that most affects the output of the model at each iteration, significantly augmenting the impact of each single injection.
The rationale behind choosing a specific position within the current window --- rather than just focusing on the type of \ac{API} --- is that single API calls cannot be strictly classified as malicious or benign. Therefore, the impact of injecting an API call may vary significantly depending on its position within the original sequence. Indeed, some positions may disrupt ``malicious'' windows, while others may have little to no effect on the final classification. Furthermore, as the attacker is only allowed to \textit{add} \ac{API} calls, it intuitively follows that the algorithm must work by displacing a certain \ac{API} within a sequence, rather than eliminating it. 

{
\begin{algorithm}[t]
\fontsize{7pt}{8pt}\selectfont
\caption{\acl{PS-FGSM}}\label{alg:advmlattack}
\begin{algorithmic}[1]
\Require $f$ (target model), $x$ (malicious sequence to perturb), $n$ (window size)
\Procedure{AdversarialAttack}{$f, x, n$}
    \State $x_{new} \gets [\hspace{0.5em}]$
    \State $r \gets 0$
    \For{window $w_j$ of size $n$ in $x$}
        \While{stop\_condition}
            \State $\mathbf{J_j} \gets J_f(w_j)[f(w_j)]$
            \State $\mathbf{P^{*}} \gets$ \Call{ComputeBestPosition} {$\mathbf{J_j}$, $r$}
            \State $w_j[\mathbf{P^{*}}] \gets \text{arg}\min_{a}||sign(w_j[0:\mathbf{P^{*}}-1] \perp a \perp w_j[\mathbf{P^{*}}:n-1]) - sign(\mathbf{J_j})||$
            \State $r \gets r + 1$
        \EndWhile
        \State $x_{new} \gets x_{new} \perp w_j$
    \EndFor
\State \Return $x_{new}$
\EndProcedure
\Procedure{ComputeBestPosition}{$\mathbf{J_j}$, $r$}
    \If{$r \mod 4 == 0$}
        \State $norm \gets \left\| \mathbf{J_j} \right\|_1$
    \Else
        \State $norm \gets \left\| \mathbf{J_j} \right\|_{-1}$
    \EndIf
    \State $\mathbf{P^{*}} \gets \text{argmax}(norm)$
\State \Return $\mathbf{P^{*}}$
\EndProcedure
\vspace{-0.1cm}
\end{algorithmic}
\end{algorithm}
}

In Algorithm~\ref{alg:advmlattack}, which showcases the \ac{PS-FGSM} algorithm for \ac{API} sequence-based malware detectors, $\bot$ represents the concatenation operation. In its basic implementation, the $stop\_condition$ is triggered either when the oracle is evaded or when the maximum number of injected API calls is reached. At each iteration, the algorithm processes a window of API calls of size $n$, which is constrained by the accepted input sequence length of the oracle model. Sequences shorter than $n$ are padded, whereas sequences longer than $n$ are truncated. Moreover, our approach includes a mechanism to shift the last API call from the current window to the subsequent one each time a new API call is injected. 
The position where we inject the additional \ac{API} call is selected by looking at the norms of the Jacobian matrix. Being \ac{API} calls discrete entities, these are generally embedded into multidimensional tensors of order $k$ by the classifier. Hence, to evaluate the impact of each \ac{API} call, we compute their norms along the second-last dimension of their embedding (i.e., assuming $k$ is the number of dimensions of the embedding, we only consider the $d_{k-1}^{th}$ dimension). In this way, the algorithm will produce a matrix of norms with one column for each \ac{API} call. To select the best candidate in the window, we then select the position with the \textit{greatest smallest norm}. In other words, we collect the smallest norm for each \ac{API} in that window and then select the greatest. We have empirically verified that this works generally better than selecting the position corresponding to the \textit{greatest absolute norm}. Nonetheless, selecting the \textit{greatest smallest norm} may lead the solution to stagnate; we combine the two norms by adopting the \textit{greatest absolute norm} every $c$ iteration, where $c$ is a small integer ($4$ in our experiments). Given $ U_i\in \mathbb{R}^{d\times n}$ and $ U_i~:=~\left[u_1,\ldots,u_{n-1},u_n\right]
    \text{s.t.}~u_j~=~\left[1, 1, \ldots, 1\right] \text{ iff } j~=~i \text{  else  } u_j~=~\left[0, 0, \ldots, 0\right]$, this process is defined as:

\vspace{-1.5\baselineskip}
\begin{align*}
    \mathbf{N^{w,r}} = \sum_i^n \mathbf{\ell_1}\left(\mathbf{J}_{emb\_api_i}\right) * U_i
\end{align*}
\vspace{-1.5\baselineskip}
\begin{align*}
    \mathbf{P_1^{w,r,*}} := \text{arg}\max_i^n\left(\min \mathbf{N}^{w,r}_i\right) , 
    \mathbf{P_2^{w,r,*}} := \text{arg}\max_i^n\left(\max \mathbf{N}^{w,r}_i\right)
\end{align*}
\vspace{-1.3\baselineskip}

In the above equations, $\mathbf{N^{w,r}}$ is the norm matrix computed for the $\mathbf{w}^{th}$ \ac{API} call window at the $\mathbf{r}^{th}$ iteration. $\mathbf{P_1^{w,r,*}}$ and $\mathbf{P_2^{w,r,*}}$ both compute the best position in the $\mathbf{w}^{th}$ window at the $\mathbf{r}^{th}$ iteration, with $\mathbf{P_1^{w,r,*}}$ selecting the one corresponding to the \textit{greatest smallest norm} whereas $\mathbf{P_2^{w,r,*}}$ selects the position where the \textit{greatest absolute norm} is found.

After determining the optimal position, the \ac{PS-FGSM} selects the \ac{API} call for injection among those that do not disrupt the original functionality of the program. Moreover, the set is restricted to API calls that are both accurately tracked by Cuckoo and employed as features by the oracle. We denote the ordered set of available \ac{API} calls as $A$. Adding API calls not belonging to $A$ would not affect the final classification, as they would be filtered out before classification in the problem space.

Given a window of \ac{API} calls $w$ and a best position $\mathbf{P^{*}}$ with respect to $w$, \ac{PS-FGSM} selects the \ac{API} call $a_{inj}\in A$ as:

\vspace{-1.5\baselineskip}
\begin{align*}
    a_{inj} = \text{arg}\min_{a\in A} ||sign(w\left[0 : \mathbf{P^{*}} - 1\right] \perp a \perp w\left[\mathbf{P^{*}}:n - 1\right]) - sign(\mathbf{J})|| \vspace{-0.2cm}
\end{align*}
\vspace{-1.3\baselineskip}

In the above equation, we compute the $\ell_1$ norm between the sign of the modified window ($w\left[0 : \mathbf{P^{*}} - 1\right] \perp a \perp w\left[\mathbf{P^{*}}:n - 1\right]$) and the Jacobian of the original one. In other words, we select the \ac{API} that most closely draws the modified window in the direction of the gradient, expressed by the Jacobian matrix $\mathbf{J}$ (similarly to~\cite{rosenberg2018generic}). 
Despite the constraints posed by the limited set of API calls --- a detail frequently overlooked in previous research
~\cite{ming2017impeding,hu2023generating,hu2017black,rosenberg2018generic,rosenberg2020query} --- Tarallo still manages to evade detection using a particularly small number of injected API calls, especially when compared to Rosenberg et al.~\cite{rosenberg2018generic}.

\mypar{Dealing with nondeterminism}
As a premise for this work, we stated that state-of-the-art attacks do not address the nondeterminism typical of real-world malware. This characteristic results in differences among behaviors observed across multiple executions, which in turn strongly impacts the effectiveness of the attack in the problem space. 
Several factors contribute to nondeterminism, including the system's state at the time of execution (e.g., available resources and network status), operating system scheduling combined with hardware and software multithreading~\cite{liu2011dthreads}, interactions with other executing processes, and evasion techniques like probabilistic control flow approaches~\cite{pawlowski2016probfuscation} and dormant periods~\cite{comparetti2010identifying}.
These factors affect the API call sequences observed during different executions of the same malware, a phenomenon that is known in the literature~\cite{polino2015jackdaw} and we encountered in our experiments. While nondeterminism has been studied in the context of dynamic analysis, to the best of our knowledge, this work is the first to address this problem in the context of evading behavioral malware detectors that rely on API call sequences.

We propose two possible strategies to address this issue: \acf{LKB} and \acf{BCO}, which maximize the attack success in case the exhibited behavior differs from the one observed by the adversary.
Both variations build upon the intuition that the success probability will increase as the confidence score associated with the malicious class decreases. Therefore, the general \ac{PS-FGSM} is modified to not interrupt the optimization process for the current window when this is classified as benign by the oracle. Instead, the window is optimized until all $n_A$ allowed \ac{API} calls are injected. The algorithm stores every prediction score obtained after each injection, along with the partial \ac{API} call sequence composed so far. After all $n_A$ sequences are generated, the algorithm then selects the one that obtained the lowest score by the oracle (i.e., the closest to the goodware label $0$). 
The \ac{LKB} variant runs the same sample $b$ times, obtaining this way up to $b$ different behaviors. It then selects the behavior that contains the most information, namely the longest sequence of \ac{API} calls. It lastly performs the variation of the \ac{PS-FGSM}.
Conversely, the \ac{BCO} builds upon the intuition that the probability of success will increase as the number of behaviors \textit{simultaneously} considered in the feature space grows: i.e., the more behaviors the attacker considers when running the \ac{PS-FGSM}, the more successful the attack will be. Specifically, the attacker considers $b$ behaviors collected from $b$ executions of the same sample. They sort them out in descending order by looking at the classification scores given by the oracle. Then they attack the most ``malicious'' sequence and propagate the solution found to all the other behaviors. Suppose the solution found by the \ac{PS-FGSM} for the first sequence is to inject API calls $\mathbf{a_r}$ before the first $a_j$ occurrence and $\mathbf{a_b}$ before the first $a_v$, so that $s_1=|a_o|a_p|a_b|a_j|a_b|a_z|a_v|$ becomes $s_1^*=|a_o|a_p|a_b|\mathbf{a_r}|a_j|a_b|a_z|\mathbf{a_b}|a_v|$. Now suppose $s2=|a_a|a_b|a_j|a_d|$ and $s_3=|a_i|a_j|a_k|a_z|a_v|$; before computing $s_2^*$ and $s_3^*$, the solution found for $s_1$ is propagated to the two sequences, obtaining in this way: $s_2'=|a_a|a_b|\mathbf{a_r}|a_j|a_d|$ and $s_3'=|a_i|\mathbf{a_r}|a_j|a_k|a_z|\mathbf{a_b}|a_v|$. The attacker then computes the solution $s_2'^*$ from $s_2'$ and propagates the results to $s_1^*$ and $s_3'$. The algorithm repeats this procedure until all the modified sequences are \textit{simultaneously} classified as benign by the oracle.

\subsection{PE Patcher}
Lastly, the PE patcher component is responsible for altering the dynamic behavior of malware samples to resemble the adversarial sequence generated by the \ac{PS-FGSM}. The PE patcher does not require access to the malware source code but is able to preserve its functionality. It supports both x86 and x86-64 PEs. 
The PE patcher identifies the assembly instructions corresponding to calls to imported APIs through heuristics (e.g., matching call and jump assembly instruction opcodes). It then modifies the arguments of those instructions to redirect them to a code snippet that implements the \textit{hijacking} logic and that was previously injected into the \ac{PE}. Indeed, our \ac{PE} patcher directly modifies the malware bytecode, in contrast to state-of-the-art solutions that rely on a wrapper as an intermediary between the malware and the DLLs. This approach makes the PE patcher arguably stealthier. 
The PE patcher can selectively choose which assembly calls to hijack and modify, while leaving other API calls of the same type unaltered. This level of granularity and flexibility enables to replicate the desired \ac{API} call sequences with high precision.

\mypar{Hijacking Logic}
When hijacking many API calls, allocating a separate handler for each is inefficient. Such an approach would require modifying extensively the \ac{PE}, compromising the attack's stealthiness. Instead, the PE patcher creates a single, adaptable code segment capable of handling the different injections. The underlying logic operates following the static directives produced by the \ac{PS-FGSM}. These directives specify which API calls to hijack and inject.
The code segment produced by the \ac{PE} patcher consists of two modules. The first one is a jump table, which acts as a trampoline for the second module. This second module contains the actual code responsible for performing both the injected and hijacked API calls. 
Upon hijacking an API call, its assembly call instruction is modified to point to a specific entry in the jump table. Each entry in this table sets up the stack with relevant information about the current execution context, such as the details of the specific API call that has been hijacked. The entire hijacking logic is placed in an additional section added at the end of the \ac{PE}. 

\mypar{Functionality Preservation}\label{ss:Functionality Preservation}
One of the main goals of the Tarallo framework is to preserve the functionality of the malware sample while altering its apparent dynamic behavior. However, formally proving that the modified program behaves exactly as the original is impossible as it can be reduced to the halting problem. Instead, Tarallo adopts a dual strategy. 

To avoid changes in its functionality, the PE patcher ensures that each API call performed by the original malware is also executed by the modified one with the same memory and register state. 
The values of caller-saved registers are stored before calling any injected \ac{API}, to later enable the reconstruction of the original stack state. In this way, Tarallo ensures that the execution of the original \ac{API} call produces the intended output, prevents subsequent injected API calls from interfering with one another, and, therefore, does not alter the original functionality. To prevent any disruption in the process flow, the PE patcher ensures that all injected \ac{API} calls are performed with valid parameters, and conform to the guidelines outlined by the Microsoft documentation~\cite{microsoft_api}. 
The PE patcher achieves this by passing the correct parameters\footnote{The list of parameters can be found at: https://github.com/necst/Tarallo/blob/main/ ChainFramework/config/config\_api\_args.py} via the stack or registers, adhering to the calling convention specified by the executable. When the API call needs a pointer as an argument, a pointer within the section added by the PE patcher is provided. If the argument is used only in read mode, the PE patcher fills the location pointed to by that pointer, with null characters or a random string of the required size~\cite{microsoft_api}. 
Conversely, if the pointer is used in write mode, a pointer to a sufficiently large area of the section, not used for other operations, is provided. Additionally, before executing any injected API call, the stack is aligned to avoid potential issues that could arise from SIMD (Single Instruction, Multiple Data) operations~\cite{simdref}.

To demonstrate the preservation of functionality between the original and modified versions, the PE patcher employs an empirical approach. The framework runs both the original and modified versions, recording their API call sequences. It then compares these sequences to determine if the modified version maintains the original behavior. If executed under the same conditions, the sequences should match, except for injected API calls in the modified version. However, replicating identical execution conditions is not always feasible due to nondeterminism. To address this issue, each malware sample is executed multiple times, both before and after the attack. From each run, Tarallo extracts a set of the executed API calls and their parameters. It then computes an invariant that represents the pre-modification behavior. This is done by computing intersections among sets of non-empty API call sequences from the original malware executions. The invariant serves as a reference point to compare against post-modification behaviors, which are derived through a set union operation of API call sequences from different executions of the modified malware. If the invariant (i.e., the original behavior) is a subset of the post-modification behavior, then Tarallo preserves the program functionality.

%% file: sections/experiments.tex
\section{Experiments}\label{sec:experiments}
The experiments aim to establish whether the \ac{PS-FGSM} is effective in deceiving state-of-the-art detectors~\cite{li2022novel,zhang2020dynamic} both in the feature and problem spaces, whether the resulting adversarial malware samples preserve their original functionality, and to measure the number of injections required to evade detection (i.e. attack overhead).
The feature space evaluation quantifies the attack ability to defeat \textit{only} the \ac{DL} classifier, which means that the object of the evaluation is solely the \ac{API} call sequence modified by the \ac{PS-FGSM}. On the other hand, the problem space evaluation concerns the ability to modify a malware sample that, upon re-execution, will display a behavior able to evade the target \ac{DL} classifier.
In brief, the following experimental evaluation aims to answer four research questions: 

\begin{tcolorbox}[width=\textwidth,boxsep=-1mm,boxrule=0.4pt]    
\fontsize{8pt}{7pt}\selectfont
    \mypar{RQ1} Is the original functionality of the malware sample preserved after applying the changes to make it evasive? (\textit{Functionality preservation})
    
    \mypar{RQ2} Can \ac{PS-FGSM} generate \ac{API} call sequences from previously detected malware that can evade state-of-the-art detectors? (\textit{Feature space experiment}).
    
    \mypar{RQ3} Can \ac{PS-FGSM} combined with either the \ac{LKB} or the \ac{BCO} effectively fool a detection system in an end-to-end manner? (\textit{White- and black-box problem space experiments}).
    
    \mypar{RQ4} Does the attack transfer to different \ac{RNN} models employing a different encoding for the \ac{API} call sequences? (\textit{Black-box problem space experiment}).
\end{tcolorbox}

\mypar{Sandbox Setup}
We employ 35 Oracle VirtualBox (v5.1.38) virtual machines running Windows 7. Each sample is run with a time limit of 300 seconds~\cite{polino2017measuring,galloro2022systematical}, a duration greater than in~\cite{rosenberg2018generic,rosenberg2020query,hariom2021adversaries}. We disabled the Cuckoo options for simulating random human interactions to ensure a controlled analysis environment and enabled deterministic human interactions, like button-clicking operations, to manage software requiring human interactions before exhibiting their behavior.
However, these options cannot encompass all scenarios, leading to software that exhibits limited behaviors in the sandbox environments. This is a recognized challenge in dynamic analysis, part of the Reverse Turing Test\cite{dynamic_evasion}.

\mypar{Datasets}
Throughout the experiments, we employ different datasets.
The first dataset, presented in~\cite{li2022novel}, includes 2000 malware samples first observed in 2019 and 2000 benign executables. The malware was sourced from VirusShare~\cite{virusshare}, and the benign samples were obtained from popular free software sources like Softonic, SourceForge, and Portableapps.
Another dataset, referenced in~\cite{zhang2020dynamic}, encompasses the features extracted from 15931 malicious samples detected in April 2017 and an additional 11856 samples from May 2017. This dataset also contains 11417 benign samples gathered in April 2017 and a further 21983 samples collected in May 2017.
We also employ Dataset n.375 from VirusShare, comprising 5474 PE files. Of these, 4917 are 32-bit \acp{PE}, and 557 are x86-64 PE files.
Finally, Dataset n.290 from VirusShare is used, consisting of 7651 PE files. Among these, 7636 are 32-bit and 15 are 64-bit, all collected in May 2017.

\mypar{Functionality Preservation (RQ1)} 
We executed the original and the modified malware five times in the Cuckoo sandbox. Comparing their behaviors, we observed functionality preservation in 89\% of the cases, slightly lower than Rosenberg et al.~\cite{rosenberg2018generic} where all modified malware maintain their functionality. However, 89\% is a lower bound. Some differences might result from nondeterminism rather than actual disruptions in functionality, a factor not treated by previous works that do not execute the original and modified malware multiple times. Additionally, Tarallo modifies the malware's bytecode directly instead of using wrappers, increasing realism but also implementation complexity.

\subsection{Feature Space Experiment (RQ2)}
The feature space experiment evaluates the performance of the \ac{PS-FGSM} against both state-of-the-art and random algorithms in deceiving the malware detector presented by Li et al.~\cite{li2022novel}.
This evaluation is conducted in a white-box setting, where both the oracle and the target model are the same as Li et al.~\cite{li2022novel}. Each adversarial sequence is generated from a malicious API call sequence extracted from the execution of a real malware sample. Each test is deemed successful if the adversarial sequence is misclassified as benign by the target detector. Additionally, we measure the number of injected API calls required to achieve misclassification. We refer to this number as the ``attack overhead,'' and we use this as an indicator of the attack efficiency.

\begin{figure}[t]
  \centering
  \includegraphics[width=0.75\linewidth]{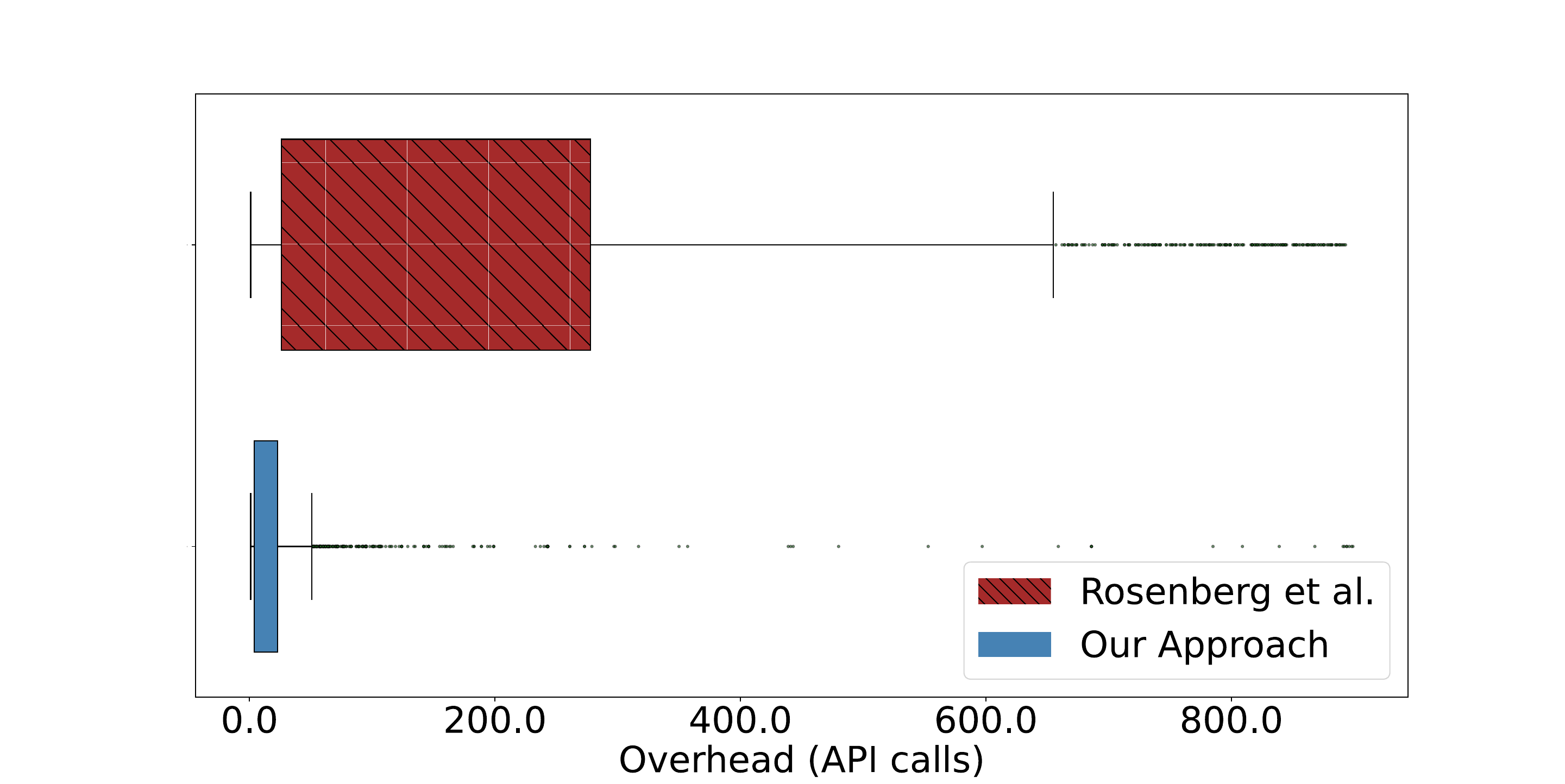}
  \caption{Box plot of the overhead distribution for the feature-level attack against Li et al.~\cite{li2022novel}, using both the proposed PS-FGSM and Rosenberg et al.~\cite{rosenberg2018generic}.}
  \label{fig:box-plot}
  \vspace{-0.5 cm}
\end{figure}

\mypar{Evaluation Result}
In this experiment, we employ the VirusShare dataset n.375 and the one used in~\cite{li2022novel}. The first one is the closest temporally to the one used to train the oracle. Using datasets from different time periods would result in a weak performance by the target detector, which would in turn introduce a bias in favor of the adversarial attack success. We also filter out all the malware samples not detected by the target model, ending up with 2708 malware samples from the VirusShare dataset and 2000 from the one in~\cite{li2022novel}.
\begin{table}[t]
    \centering
    \caption{Effectiveness scores for the \textbf{feature-level} attack against Li et al.~\cite{li2022novel} using the proposed PS-FGSM, Rosenberg et al.\cite{rosenberg2018generic}, and a random approach, evaluated with both the VirusShare Dataset No. 375 and the Li et al. Dataset~\cite{li2022novel}.}
    \label{tab:combined-datasets}
    \fontsize{7.5pt}{8pt}\selectfont
    \begin{tabular*}{\textwidth}{@{\extracolsep{\fill}}ccccccc}

        \toprule
        & \multicolumn{3}{c}{VirusShare Dataset n.375} & \multicolumn{3}{c}{Li et al. Dataset~\cite{li2022novel}} \\
        \cmidrule(lr){2-4} \cmidrule(lr){5-7}
        Overhead limit & 20\% & 50\% & 70\% & 20\% & 50\% & 70\% \\
        \midrule
        PS-FGSM & \textbf{0.9742} & \textbf{0.9841} & \textbf{0.9863} & \textbf{0.9815} & \textbf{0.9920} & \textbf{0.9925} \\
        Rosenberg et al.\cite{rosenberg2018generic} & 0.6492 & 0.8556 & 0.9261 & 0.5696 & 0.8178 & 0.9044 \\
        Random Approach  & 0.2670 & 0.3674 & 0.7020 & 0.2638 & 0.3634 & 0.7192 \\
        \bottomrule
            \vspace{-0.5 cm}

    \end{tabular*}

\end{table}
In Table~\ref{tab:combined-datasets}, we present the results obtained with our \ac{PS-FGSM}, the approach devised in~\cite{rosenberg2018generic}, and an attack consisting in randomly injecting \ac{API} calls. 
Both \ac{PS-FGSM} and the approach in~\cite{rosenberg2018generic} demonstrate almost 100\% effectiveness when not imposing strict limits on the attack overhead. However, when we consider a more reasonable attack overhead, the \ac{PS-FGSM} significantly outperforms the approach in~\cite{rosenberg2018generic}. The box plot in Figure~\ref{fig:box-plot} provides a visual representation of the attack overhead by \ac{PS-FGSM} and the approach in~\cite{rosenberg2018generic}. 
In the VirusShare dataset, the average overhead for \ac{PS-FGSM} is 27, indicating that, on average, we have to inject only 27 additional API calls to evade detection. In contrast, the approach in~\cite{rosenberg2018generic} has an average overhead of 202. The same happens for dataset in~\cite{li2022novel}, where the average overhead for the \ac{PS-FGSM} is 30, while for~\cite{rosenberg2018generic} is 252. Furthermore, when looking at the median overhead, we observe that half of the evaluated samples require 12 or fewer API calls to achieve misclassification when employing the \ac{PS-FGSM}. In contrast, the median overhead for the approach in~\cite{rosenberg2018generic} is 110, indicating that the majority of samples require a much higher number of injected API calls. Similarly, the median overhead of PS-FGSM is 13 for the dataset in~\cite{li2022novel}, while the approach in~\cite{rosenberg2018generic} has an average overhead of 165.

\subsection{White-Box Problem Space Experiment (RQ3)}
\label{wbexperiment}
This experiment shows the capability of Tarallo to produce a modified malware with an evasive dynamic behavior. These settings present additional challenges compared to the feature space attack; e.g., the solution computed in the feature space may include calls to \ac{API} that are not imported by the original malware.  Moreover, it explicitly addresses the nondeterminism in the execution of malware samples in a sandbox, which results in different \ac{API} call sequences during different executions of the same malware sample, a problem not addressed in previous works such as~\cite{rosenberg2018generic}. To explicitly tackle the complexity of the problem space settings, we employ the \ac{LKB} and the \ac{BCO} strategies, both based on the \ac{PS-FGSM}.
We consider the attack successful if and only if re-executing the malicious sample produces an \ac{API} call sequence that is misclassified as benign by the target model. 
In the white-box tests, we use the malware detector described by Li et al. in~\cite{li2022novel} as both the oracle and the target model.

\mypar{Evaluation Result}
The VirusShare dataset n.375 is employed in this experiment; in particular, we keep only executables classified as malicious by the target model. Moreover, we exclude all malware with API call sequences shorter than a threshold set to 15 \acp{API}, ensuring that only malware with significant recorded dynamic behavior is considered for modification. For each malware sample, we define an ``arsenal'' of API calls available for injection, which comprises the intersection of the \acp{API} imported by the malware, the \acp{API} considered safe to inject without disrupting its normal execution, and the \acp{API} recorded by Cuckoo. The results are evaluated using various arsenal size ranges to measure their impact on the attack performance. 
To evaluate the attack's effectiveness, we run each malware sample five times in Cuckoo. Then, after applying the \ac{LKB} and \ac{BCO} strategies, each modified sample is executed five more times per strategy. The resulting API call sequences are submitted for evaluation.
In Table~\ref{tab:end-to-end-test}, we present the results of this evaluation, detailing different arsenal ranges and overhead limits. The arsenal ranges are expressed as the percentage of the total number of API calls recorded by Cuckoo, which is just over 300. For example, a 2\% value corresponds to approximately 6 API calls. We compute the attack effectiveness as $Evasive Executions / Total Executions$.

\begin{table}[t]
    \centering
    \fontsize{7.5pt}{8pt}\selectfont
    \caption{\acf{LKB} and \acf{BCO} \textbf{white-box} attacks overhead and effectiveness in \textbf{problem space} against Li et al.~\cite{li2022novel}.}
    \label{tab:end-to-end-test}
    \begin{tabular*}{\textwidth}{@{\extracolsep{\fill}}lccccccccc}
        \toprule
        & Overhead Limit & \multicolumn{3}{c}{5\%} & \multicolumn{3}{c}{20\%} & \multicolumn{2}{c}{120\%}\\
        \cmidrule(lr){3-5} \cmidrule(lr){6-8} \cmidrule(lr){9-10}
        & Arsenal size & (2\%, 3\%] & (2\%, 6\%] & Any & (2\%, 3\%] & (2\%, 6\%] & Any & Any & \\
        \midrule
        \multirow{4}{*}{\textbf{\ac{LKB}}} & Total Executions & 73 & 165 & 416 & 115 & 267 & 800 & 1414 & \\
        & Evasive Executions & 72 & 129 & 249 & 98 & 189 & 455 & 806 & \\
        & Avg Injected APIs & $\approx 19$ & $\approx 19$ & $\approx 22$ & $\approx 56$ & $\approx 56$ & $\approx 74$ & $\approx 406$ & \\
        & Attack Effectiveness & \textbf{0.9863} & 0.7818 & 0.5986 & 0.8522 & 0.7079 & 0.5688 & 0.5700 & \\
        \midrule
        \multirow{4}{*}{\textbf{\ac{BCO}}} & Total Executions & 90 & 219 & 723 & 100 & 278 & 1120 & 1571 & \\
        & Evasive Executions & 80 & 196 & 629 & 90 & 252 & 837 & 1018 & \\
        & Avg Injected APIs & $\approx 15$ & $\approx 17$ & $\approx 18$ & $\approx 25$ & $\approx 40$ & $\approx 56$ & $\approx 219$ & \\
        & Attack Effectiveness & 0.8889 & \textbf{0.8950} & \textbf{0.8700} & \textbf{0.9000} & \textbf{0.9065} & \textbf{0.7473} & \textbf{0.6480} & \\
        \bottomrule
    \end{tabular*}
    \vspace{-0.5 cm}
\end{table}

The problem space constraints have an impact on the attack effectiveness to some extent. However, despite the complex settings, the modified malware still manages to exhibit an evasive apparent dynamic behavior that successfully deceives the white-box target model. For the \ac{LKB} strategy, we find that the attack effectiveness score decreases as the arsenal size and the overhead limit increase. A larger arsenal size implies a greater variety of API calls available to the malware, which implies that the problem space is broader and, therefore, more challenging from the attacker's perspective, as it amplifies the impact of nondeterminism on the attack performance. As a result, when the malware is re-executed post-attack, the resulting \ac{API} call sequence may significantly deviate from the original, including differences in API calls used. Moreover, increased overhead generally results in a more fragile attack. Specifically, a high number of injected \ac{API} calls has an increased chance that variations in the sequence of API calls due to nondeterminism will shift these injections, rendering them ineffective. The \ac{BCO} strategy shows a more stable attack effectiveness than the \ac{LKB} strategy, underscoring its superior ability to navigate the breadth of the problem space and handle variations in executed behaviors due to nondeterminism. Moreover, the \ac{BCO} strategy generally outperforms the \ac{LKB} strategy, supporting the soundness of its underlying rationale. The only instance where the \ac{LKB} strategy surpasses \ac{BCO} is in scenarios with the most limited overhead and arsenal size. This is expected as it is the scenario with a narrower problem space, where nondeterminism has minimal impact. This observation further reinforces the link between the \ac{BCO} strategy's effectiveness and nondeterminism.

Additionally, we analyze the ability of the \ac{BCO} and \ac{LKB} strategies to generate adversarial API call sequences in a constrained environment. We use a subset of the VirusShare dataset no. 375, limiting to a maximum of 800 injected API calls per window and an overall overhead of 120\%. Interestingly, the \ac{BCO} and \ac{LKB} strategies are successful with different malware samples. In our tests, only about 15\% of the samples with an adversarial sequence are targeted by both strategies. Furthermore, the malware samples effectively attacked by the \ac{BCO} strategy have an average pre-attack sequence length that is roughly 25\% shorter than those targeted by the \ac{LKB} strategy.
The \ac{LKB} strategy succeeds with the most informative (i.e., longest) sequence across different executions of the same malware sample, showing greater effectiveness in creating adversarial sequences with longer available sequences. Conversely, the \ac{BCO} strategy derives information from all executions of a sample. Consequently, shorter sequences in each run, being less informative, fit within the set overhead limit, making \ac{BCO} effective in these cases. Notably, unlike the \ac{LKB} strategy, the \ac{BCO} strategy’s suggested injections are meant for different executions of the same sample, allowing for higher overhead limits compared to \ac{LKB}. With increased overhead, \ac{BCO} can attack more samples, including some previously only vulnerable to \ac{LKB}, although these do not completely overlap.

\subsection{Black-Box Problem Space Experiment (RQ3, RQ4)}
The black-box, problem space test follows a procedure similar to the white-box one, with a key difference. In this test, the final API call sequences extracted from the modified malware are not evaluated against the model by Li et al.~\cite{li2022novel}, which serves as the oracle for the \ac{PS-FGSM} attack. Instead, these sequences are tested against a different model, unknown to the attacker. The target model for this black-box test is the \ac{DL} framework proposed by Zhang et al.~\cite{zhang2020dynamic}. This model presents a significant challenge as it takes into account the arguments of API calls, which are not explicitly manipulated by the AML attack. Instead, the arguments for the injected API calls are defined by the PE patcher to be valid and ensure they do not disrupt the malware's normal execution.

\begin{table}[t]
\centering
\caption{Effectiveness scores of \acf{LKB} and \acf{BCO} approaches in \textbf{black-box}, \textbf{problem space} settings against Zhang et al.\cite{zhang2020dynamic}—evaluated across different prediction thresholds, \acf{TPR}, and \acf{FPR}—using the model by Li et al.\cite{li2022novel} as an oracle.}
\label{tab:EE-BB}
\fontsize{7.5pt}{8pt}\selectfont
\begin{tabular*}{\textwidth}{@{\extracolsep{\fill}}ccccccccc}
\toprule
Threshold & TPR & FPR & \multicolumn{2}{c}{\begin{tabular}[c]{@{}c@{}}Available\\ Executions\end{tabular}} & \multicolumn{2}{c}{\begin{tabular}[c]{@{}c@{}}Evading\\ Executions\end{tabular}} & \multicolumn{2}{c}{\begin{tabular}[c]{@{}c@{}}Attack Effectiveness\end{tabular}} \\ 

\cmidrule(lr){4-5} \cmidrule(lr){6-7} \cmidrule(lr){8-9}
 &  &  & \ac{LKB} & \ac{BCO} & \ac{LKB} & \ac{BCO}  & \ac{LKB} & \ac{BCO}  \\

\midrule
0.9683                       & 0.992 & 0.325  &  309 & 161    & 154 & 82 & 0.4984    & \textbf{0.5093}      \\
0.9776                       & 0.989 & 0.300  &  263 & 146   & 130  & 71 & \textbf{0.4943}    & 0.4863    \\
0.9868                       & 0.986 & 0.275  &  225 & 131   & 115  & 75 & 0.5111 & \textbf{0.5725}          \\
0.9951                       & 0.978 & 0.225  &  162 & 86   & 79  & 51 & 0.4877   & \textbf{0.5930}        \\
0.9981                       & 0.971 & 0.178  &  141 & 82   & 69  & 52 & 0.4894   & \textbf{0.6341}        \\
0.9988                        & 0.968 & 0.150  & 132 & 68    & 67 & 39 & 0.5076   & \textbf{0.5735}            \\
0.9996                        & 0.958 & 0.100  & 115 & 57  & 54  & 33 & 0.4696   & \textbf{0.5789}          \\
\bottomrule
\end{tabular*}
\vspace{-0.8 cm}
\end{table}

\mypar{Evaluation Result}
We train the black-box model using the dataset and code provided in~\cite{zhang2020dynamic}. We follow the recommendation from the original work to set the prediction threshold corresponding to a \ac{FPR} of 0.1\%. However, we encountered challenges with this configuration as the model is unable to accurately identify the original malware samples from both dataset n.375 and dataset n.290 from VirusShare. This suggests that the chosen threshold and configuration may not be suitable for our specific purposes and dataset. Hence, we select threshold values that correspond to high \acp{TPR}, rather than focusing on a specific \ac{FPR}. Although this strategy results in higher \acp{FPR}, it allows us to assess the effectiveness of the Tarallo framework in fooling robust machine learning systems.
 By testing our method with different threshold values, we can demonstrate how this parameter influences the effectiveness of our attack and provide a comprehensive validation of the experimental evaluation.
In our tests, we run the original malware samples from the VirusShare dataset n.375 through the Cuckoo sandbox, recording their API call sequences. We repeat this process five times to collect a sufficient number of executions. For each threshold value considered during the test, we select only the malware samples that are classified as malicious by the black-box model in \textit{all} the executions. We attack each sample with both the \ac{LKB} and \ac{BCO} attack strategies and then run the modified versions of these malware executables through the Cuckoo sandbox five times for each strategy, recording their evading API call sequences. The results of these experiments are summarized in Table~\ref{tab:EE-BB}. The attack effectiveness is instead computed as $Evading Executions / Total Executions$.
    Restricting the target model to a lower \ac{FPR} —though still above the value reported in the original work— by increasing the classification threshold, leads to fewer correctly classified malware samples (i.e., a lower \ac{TPR}), particularly those the model is most confident about. This makes attacks on these samples harder for both the \ac{LKB} and \ac{BCO}, reducing the effectiveness score compared to other threshold settings. Lowering the threshold results in an increase in the \ac{FPR}, which may be less realistic, but it also enhances the \ac{TPR}. This improvement boosts the model's malware detection capabilities, more accurately mirroring real-world conditions. Under these conditions, both attacks tend to exhibit improved performance.
However, when we further increase both the \ac{TPR} and the \ac{FPR}, the attack effectiveness score starts to decline again. This decrease occurs because the model begins to incorrectly identify a large number of benign behaviors as malicious, leading to a misclassification of the injection in the original API call sequence as harmful.
The tests underscore the need to balance \ac{TPR} and \ac{FPR} in detection system design, requiring careful domain-specific consideration to optimize accuracy while reducing evasion attack risks. Additionally, the consistently better performance of the \ac{BCO} strategy confirms that considering multiple sequences during the attack results in more resilient and general adversarial sequences.

%% file: sections/conclusions.tex
\section{Conclusions}\label{sec:conclusions}

In this paper, we introduced Tarallo, an end-to-end framework designed to modify the apparent dynamic behavior of malware to deceive \ac{ML} detectors that employ \ac{API} call sequences as features.
Our first major contribution is a novel \ac{AML} attack that targets discrete and sequential data. Moreover, we explicitly addressed the problem of nondeterministic execution in a sandbox, which makes the design of problem space attacks particularly challenging. Lastly, we introduced a new approach for modifying the run-time behavior of malicious software that does not require access to the source code. 
The preliminary experimental evaluation in a sandboxed environment and against two \ac{RNN}-based malware detectors, showed that our novel \ac{AML} algorithm, namely the \ac{PS-FGSM}, outperforms current state-of-the-art algorithms in terms of both effectiveness and efficiency. In both feature and problem spaces, our algorithm achieves up to 99\% effectiveness, while also minimizing overhead as measured by the number of injected APIs.
Although we showed that obfuscating the behavior of existing malware is a nontrivial problem and, therefore, that current antimalware solutions are very robust, we also demonstrated that such attacks are in fact possible. 

Future work will focus on assessing the effectiveness of Tarallo against different malware detectors, especially those not based on \ac{RNN}. Moreover, we will explore methodologies that allow defenders to remove non-operational APIs from API call sequences, enhancing the overall performance of the detection systems. We also intend to employ the \ac{PS-FGSM} to build an adversarial training module to strengthen state-of-the-art \ac{DL}-based detectors.